\documentclass[lettersize,journal]{IEEEtran}
\usepackage{amsmath,amsfonts}
\usepackage{algorithmic}
\usepackage{algorithm}
\usepackage{array}
\usepackage[caption=false,font=normalsize,labelfont=sf,textfont=sf]{subfig}
\usepackage{textcomp}
\usepackage{stfloats}
\usepackage{url}
\usepackage{verbatim}
\usepackage{graphicx}
\usepackage{cite}
\begin{document}

\title{Density-encoded line integral convolution: polarisation optical~axis~tractography using centroidal~Voronoi~tessellation}

\author{Darven~Murali~Tharan, Marco~Bonesi, Daniel~Everett, Cushla~McGoverin, Sue~McGlashan, Ashvin~Thambyah, Frédérique~Vanholsbeeck

\thanks{D.~Murali~Tharan, M.~Bonesi, C.~McGoverin and F.~Vanholsbeeck are with the Department of Physics, The University of Auckland, Auckland, New Zealand.}

\thanks{D.~Murali~Tharan, M.~Bonesi, D.~Everett, C.~McGoverin and F.~Vanholsbeeck are with the Dodd-Walls Centre for photonic and quantum technologies, New Zealand.}

\thanks{D.~Everett and A.~Thambyah are with the Department of Chemical and Materials Engineering, The University of Auckland, Auckland, New Zealand.}

\thanks{S. McGlashan is with the Department of Anatomy and Medical Imaging, The University of Auckland, Auckland, New Zealand.}

}

\maketitle

\begin{abstract}
Visualising complex polarimetry optical axis fields is challenging. We introduce density-encoded line integral convolution (DELIC), a novel approach that builds on the classic line integral convolution algorithm by incorporating the principles of centroidal Voronoi tessellation, enabling clearer and more interpretable representations of complex optical axis fields. 
\end{abstract}

\begin{IEEEkeywords}
Vector field visualisation, scientific visualisation, biomedical imaging, tractography, polarimetry, centroidal Voronoi tessellation.
\end{IEEEkeywords}

\section{Introduction}
\label{sect:intro}  
\IEEEPARstart{V}{ector} field visualisation plays a critical role in scientific research, aiding in the analysis of complex flow patterns. Such tools are widely used in biomedical imaging, particularly in tractography, where the structural pathways of neuronal axons, muscle fibres, and collagen are mapped. Biomedical polarimetry, the study of how tissue alters the polarisation state of light~\cite{ghosh2011tissue}, has become an essential tool for tractography in birefringent tissues~\cite{mollink2017evaluating, yao2020high, palm2010towards}. Many biological tissues, like those composed of collagen or axons, consist of highly organised fibrous structures that exhibit birefringence, altering the polarisation state of light. Polarimetry enables measurement of two key properties: birefringence strength, which reflects fibrous density and organisation, and the optical axis, indicating the orientation of these structures.

In polarimetry-based tractography, methods such as streamlines and glyphs are commonly used to visualise fibrous orientation. These methods can clutter visualisations in dense or complex fields, making interpretation challenging~\cite{tharan2025ConfProceedings}. However, these methods remain popular due to their simplicity and widespread availability, such as MATLAB’s \texttt{stream2}, Python’s \texttt{streamplot}, and R’s \texttt{geom\_path}. Line integral convolution (LIC) provides a more sophisticated approach by convolving a noise texture along streamlines, offering a clutter-free and accurate representation of the fine details in complex vector fields~\cite{lic}. Despite its advantages, LIC has seen limited use in polarimetry-based tractography. One reason for this could be the high density and high-frequency content of LIC images, which may overwhelm and become difficult to interpret~\cite{olic}. 

In this letter, we introduce density-encoded line integral convolution (DELIC), a novel algorithm specifically designed for visualising birefringent tissue inspired by the traditional LIC technique. Unlike LIC, which uses a noise texture that spans the entire domain, DELIC employs a sparse noise texture. This sparsity helps address the issue of LIC’s dense and high frequency content, which can make interpretation difficult~\cite{olic}. While reducing texture density may seem like a loss of detail compared to LIC, DELIC compensates by selectively concentrating texture density in regions of interest, such as areas of high birefringence, where the tissue structural information is assumed most significant. By modulating texture density based on birefringence strength, DELIC enhances the clarity of critical regions while simplifying less important areas, reducing visual clutter. This targeted approach preserves essential structural details, offering clearer and more interpretable visualisations with minimal loss of information.

We demonstrate the effectiveness of DELIC by direct comparison with LIC using polarimetry data acquired through a custom-built polarisation-sensitive optical coherence tomography (PS-OCT) system. Optical Coherence Tomography (OCT) is non-invasive imaging technique that is the optical equivalent to ultrasonography, capable of generating cross-sectional (sagittal plane) images of biological tissue~\cite{drexler2008optical}. PS-OCT is a functional extension of this technique that allows measurement of tissue birefringence and the optical axis of birefringence~\cite{psoct,baumann2017polarization}. 

Our results highlight how DELIC enhances the visualisation of fibrillar architecture, offering improved clarity compared to LIC. This approach represents a significant advancement in polarimetry visualisation, offering a powerful and effective tool for the clear and accurate interpretation of polarimetry data. DELIC has the potential to be applied both in scientific research and clinical settings, enabling improved \emph{in vivo} analysis and exploration of birefringent tissues.

\section{Methods}
\subsection{Line Integral Convolution (LIC)}

Consider a vector field $\mathbf{V}(\mathbf{r})$ over a spatial domain \( \Omega \subset \mathbb{R}^d \), where $\mathbf{r}$ represents a position. Streamlines are computed by integrating the vector field from each point $\mathbf{r}$, producing paths that follow the direction of the field. For each point, the path is parametrised as $\mathbf{s}(t; \mathbf{r})$, where $t$ is a parameter along the path:
\begin{equation}
\mathbf{s}(t; \mathbf{r}) = \mathbf{r} + \text{sgn}(t) \int_{0}^{|t|} \mathbf{V}(\mathbf{s}(\tau; \mathbf{r})) \, d\tau,
\label{Methods:Equation:Trajectory}
\end{equation}
where \(\text{sgn}(t) \) is the signum function. In this work, we use the 4th-order Runge-Kutta (RK4) method for numerical integration.

A randomly generated noise texture, \(I_{n}(\mathbf{r})\), is then convolved along the streamline using a kernel $k(t)$, which in this work is a Gaussian window. The LIC value at a point $\mathbf{r}$, denoted as $I_{LIC}(\mathbf{r})$, is computed as:
\begin{equation}
I_{LIC}(\mathbf{r}) = \frac{1}{K} \int_{-T}^{T} I_n(\mathbf{s}(t; \mathbf{r})) k(t) \, dt,
\label{Methods:Equation:LIC}
\end{equation}

\noindent where \(K = \int_{-T}^{T} k(t) \, dt\) and \(T\) is the convolution period.

\subsection{Density encoded line integral convolution}
\label{sect:DELIC}
While traditional LIC initialises with a random noise texture that fills the entire domain, \(\Omega\), DELIC uses a sparse texture field, where values are assigned only to a select set of seed points scattered across the domain. The rest of the domain remains empty or zero-valued. This reduces clutter and improves clarity in the resultant tractogram, making the vector field direction easier to interpret~\cite{olic}. By modulating seed point density based on an underlying scalar field, DELIC emphasises regions of interest such as areas of \textbf{high birefringence strength}, ensuring that key features are highlighted without losing detail. Moving forward, unless stated otherwise, we denote the birefringence scalar field used to generate a DELIC tractogram as the encoding field, \(\rho(\mathbf{x})\). 

Below, we start by giving a brief summary of the entire DELIC processing pipeline which leads to an in-depth description in the subsequent sections. 

\begin{itemize}
    \item \textbf{Extract and Process Data:} Retrieve birefringence strength (Fig.~\ref{fig:methods:CVT} A) and optical axis orientation data (Fig.~\ref{fig:results:Experimental} A) and define a vector field of the optical axis.
    
    \item \textbf{Seed Point Generation:} Normalize birefringence to define a probability density function (PDF), then apply rejection sampling to generate seed points reflecting the PDF (Fig.~\ref{fig:methods:CVT} B), with a minimum distance of \(d_{\mbox{sep}}\) between each seed.
    
    \item \textbf{Seed Point Refinement:} Refine the initial seed points to better reflect the PDF. (Fig.~\ref{fig:methods:CVT} C).
    
    \item \textbf{Texture generation:} Assign random grey values to the refined seed points using Sobol quasi-random generation~\cite{sobol} to create a texture image (Fig.~\ref{fig:methods:CVT} D).

    \item \textbf{LIC Generation:} Perform LIC on this texture (Fig.~\ref{fig:results:Experimental} D). Optionally, apply unsharp masking to enhance sharpness and histogram stretching to improve contrast. Colour mapping based on the optical axis orientation may also be applied.
    
\end{itemize}

\subsubsection{DELIC initialisation}
\label{Delic Init}
\begin{figure}[t]
    \centering
    \includegraphics[width=1\linewidth]{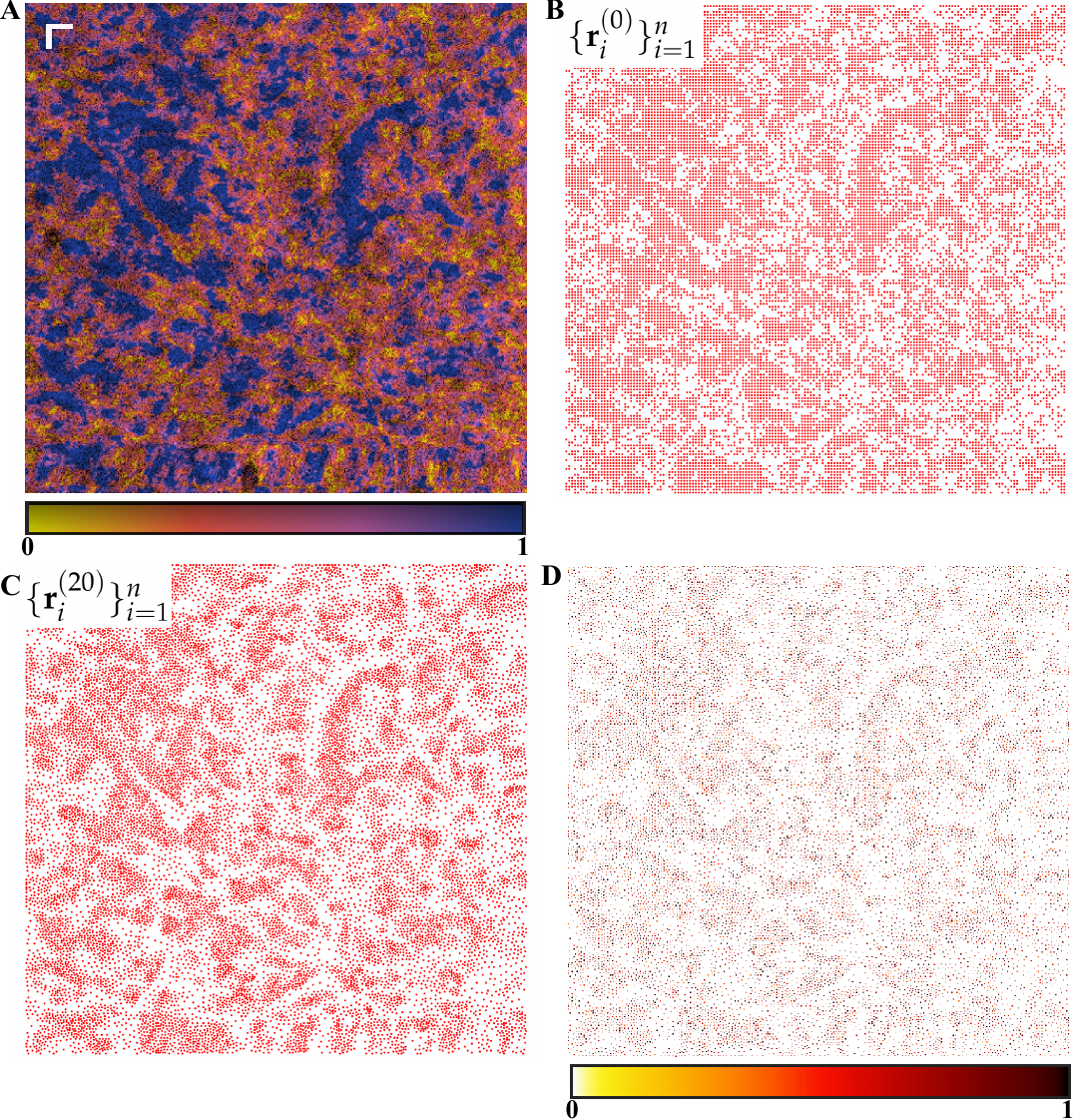}
    \caption{(A) Normalised birefringence strength image from skin. Vertical and horizontal scale bars are 500~$\mu$m long. (B) Initial seed point distribution for DELIC, generated from (A). The process begins with a grid of equidistant seed points, ensuring a minimum separation of \(d_{\text{sep}}\) between them. Each seed is then assigned a random value between 0 and 1. Finally, seeds with values lower than the corresponding normalized birefringence field are discarded. (C) Resulting seed placement after 20 iterations of Lloyds algorithm initialised by the distribution in (B). (D) Texture generated by assigning pseudo-random values, between 0.35 to 1, to the seed distribution shown in (C).}
    \label{fig:methods:CVT}
\end{figure}

Strategic placement of seed points is required for DELIC to ensure accurate representation of the encoding field. We begin with an initial point distribution generated via basic rejection sampling of the encoding field, approximating the desired density function (Fig.~\ref{fig:methods:CVT} B). We enforce a minimum distance of \(d_{\mbox{sep}}\) between each seed. This is the primary method for controlling the global density of seeds for the algorithm. This initial distribution of points, denoted \( \{\mathbf{r}_i^{(0)}\}_{i=1}^n \), serves as the starting point for further refinement. Next, we utilise the principles of \textbf{centroidal Voronoi tessellation (CVT)}~\cite{cvt} to refine \( \{\mathbf{r}_i^{(0)}\}_{i=1}^n \) so that it more accurately describes the encoding field (Fig.~\ref{fig:methods:CVT} C). 

\subsubsection{Seed refinement using centroidal Voronoi tessellation}

Given a set of points \( \{ \mathbf{r}_i\}_{i=1}^n \) within \( \Omega \), the Voronoi cell \( V_i \) associated with point \(  \mathbf{r}_i \) is defined as:

\begin{equation}
V_i = \{ \mathbf{x} \in \Omega \mid \| \mathbf{x} - \mathbf{r}_i \| < \| \mathbf{x} - \mathbf{r}_j \| \forall j \neq i \}
\end{equation}

\noindent where \(\| \| \) is the Euclidean norm in \(\mathbb{R}^d \). \( V_i \) represents the region in \(\Omega\) that is closer to  \(  \mathbf{r}_i \) than any other \(  \mathbf{r}_j \)  and satisfies:

\[
V_i \cap V_j = \emptyset \quad \forall i \neq j \hspace{0.3cm}  \& \hspace{0.3cm}   \bigcup_{i=1}^n V_i = \Omega.
\]

\noindent The set \( \{ V_i\}_{i=1}^n \) is the Voronoi tesselation of \(\Omega\) and the set \( \{ \mathbf{r}_i\}_{i=1}^n \) are the seeds or generators of the tesselation. The tesselation results in a partitioning of \(\Omega\) into \(n\) non-overlapping regions.

To generate a CVT from \( \{\mathbf{r}_i^{(0)}\}_{i=1}^n \), we minimize the following cost function:
\[
\min_{\{\mathbf{r}_i\}} \sum_{i=1}^n \int_{V_i} \| \mathbf{x} - \mathbf{r}_i \|^2 \rho(\mathbf{x}) \, d\mathbf{x}.
\]

\noindent where \( \rho(\mathbf{x}) \) is the encoding field defined by the normalised sample birefringence. The cost function represents the \( \rho(\mathbf{x}) \) weighted sum of the squared distances of all points in \( \Omega \) to their respective seed points. When \( \rho(\mathbf{x}) \)  is non-uniform, the spacing of the seed points is modulated by the encoding field. Seed points will be more densely packed in regions where \( \rho(\mathbf{x})\) is higher, reflecting the higher "importance" or "weight" of those regions.

This cost-function is minimised when each \( \mathbf{r}_i \) is at the mass centroid of their respective Voronoi cells or in other words, the Voronoi tessellation where the seed of each cell is also its centroid is the solution to the minimisation problem. A Voronoi tessellation satisfying this criterion is defined as a CVT.

To compute a CVT, the classic Lloyd's algorithm is employed~\cite{lloyds}, initialised by \( \{\mathbf{r}_i^{(0)}\}_{i=1}^n \) (Fig.~\ref{fig:methods:CVT} B). This method iteratively minimises the cost function and is computed as follows:

\begin{algorithm}[H]
\caption{Lloyd's Algorithm for CVT}
\label{alg:lloyd}
\begin{algorithmic}[1]
\STATE \textbf{Input:} Initial seed points \( \{\mathbf{r}_i^{(0)}\}_{i=1}^n \), density function \( \rho(\mathbf{x}) \)
\STATE \textbf{Output:} Updated seed points forming the CVT
\REPEAT
    \FOR{each point \( \mathbf{r}_i \)}
        \STATE Compute the Voronoi cell \( V_i \)
        \STATE Update \( \mathbf{r}_i \) using mass centroid:
        \[
        \mathbf{r}_i^{(k+1)} = \frac{\int_{V_i^{(k)}} \mathbf{x} \rho(\mathbf{x}) \, d\mathbf{x}}{\int_{V_i^{(k)}} \rho(\mathbf{x}) \, d\mathbf{x}}
        \]
    \ENDFOR
\UNTIL{Convergence criteria met}
\end{algorithmic}
\end{algorithm}

\subsubsection{Convergence criteria}

In the context of DELIC, achieving a perfectly optimized distribution of seed points is not a strict requirement; rather, the focus is on attaining a visually adequate result. In our work, we found 20 iterations of Lloyd's algorithm to be sufficient to achieve a satisfactory distribution (Fig.~\ref{fig:methods:CVT} C). While additional iterations can further refine the distribution, the processing time increases, and the improvements become increasingly marginal. Instead of initializing with an equidistant grid of points, we used rejection sampling based on the encoding field as discussed in section~\ref{Delic Init}. This method provides a starting distribution that is already closely aligned with the final expected distribution, thereby reducing the number of iterations required to refine this initial distribution.

\subsubsection{Line integral convolution step}

Once a satisfactory seed distribution is achieved (Fig.~\ref{fig:methods:CVT} C), a texture image is generated by assigning pseudo-random values between 0.35 and 1 to each seed (Fig.~\ref{fig:methods:CVT} D), utilising Sobol sequences. The lower bound of 0.35 is chosen to prevent excessively dark regions in the final DELIC image. The RK4 method is then used to compute the trajectories of the seeds (Equation.~\ref{Methods:Equation:Trajectory}) within the optical axis vector field (Fig.~\ref{fig:results:Experimental} A). Finally, the texture image is convolved along these trajectories (Equation.~\ref{Methods:Equation:LIC}), producing the final DELIC visualisation (Fig.~\ref{fig:results:Numerical} B and Fig.~\ref{fig:results:Experimental} C). We also ignore $I_n(\mathbf{s}(t; \mathbf{r}))$  values that are below a set threshold (see Equation.~\ref{Methods:Equation:LIC}) to prevent excessive darkening of the DELIC image.

\subsection{Numerical validation}

To numerically validate the ability of DELIC to visualise a vector field with density modulated by some scalar field, we utilise an artificially generated optical axis vector field.

This vector field describes a sinusoidal in the vertical direction and a logarithmic growth in the horizontal direction (Fig.~\ref{fig:results:Numerical}~B). We define an artificial birefringence scalar field as the DELIC encoding filed by utilising the normalised magnitude of this vector field. The normalised magnitude of the vector field is shown in Fig.~\ref{fig:results:Numerical} A.

\subsection{Experimental validation}

Polarimetry data was acquired using a custom-built PS-OCT system, which we have previously described~\cite{tharan2025ConfProceedings}. For detailed descriptions of PS-OCT, we refer the reader to De Boer \emph{et al.}~\cite{psoct} and for a description of the algorithms we used to reconstruct polarimetry data we refer the reader to Li \emph{et al.}~\cite{li2018robust} and Villiger \emph{et al.}~\cite{villiger2016deep} PS-OCT volume scans of human dorsal hand skin was acquired non-invasively from the author. An en face plane within the volume, approximately 400~$\mu$m below the surface of the skin, was used for our demonstration. Human skin has a complex architecture, with collagen fibrils forming a complex mesh, while wrapping around structures like sweat glands~\cite{li2020vivo}, presenting an excellent environment to assess DELIC and compare it with LIC. 

\section{Results and Discussion}

\subsection{Numerical validation}

\begin{figure}
    \centering
    \includegraphics[width=1\linewidth]{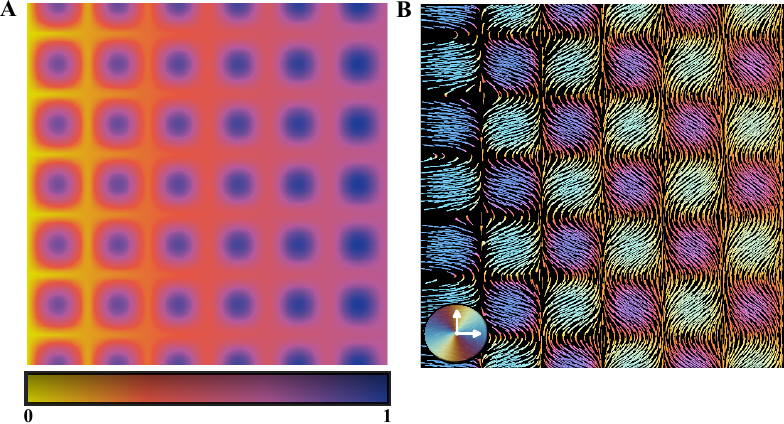}
    \caption{(A) Artificially generated encoding field from normalised vector field magnitude. (B) DELIC tractogram of the artificial vector field with \(d_{\mbox{sep}}\) = 5 pixels, integration step of 0.1 pixels and integration length of 50 steps. We apply a colourmap to the DELIC with the colours associated with the orientation of the artificial vector field (Note orientation colour wheel on the bottom left of the image).}
    \label{fig:results:Numerical}
\end{figure}

Our numerical results are presented in Fig.~\ref{fig:results:Numerical}. The DELIC tractogram (Fig.~\ref{fig:results:Numerical} B) accurately represents the orientation of the vector field. The visual orientation of the tractogram streamlines corresponds precisely to the colour coding applied. This consistent alignment between the visual orientation and the color coding confirms that the tractogram correctly encodes the vector field's orientation. Furthermore, regions with larger encoding field magnitude (Fig.~\ref{fig:results:Numerical} A) are depicted with greater tractogram density, successfully reflecting the scalar field encoding.

\subsection{Experimental validation}

\begin{figure}[H]
    \centering
    \includegraphics[width=1\linewidth]{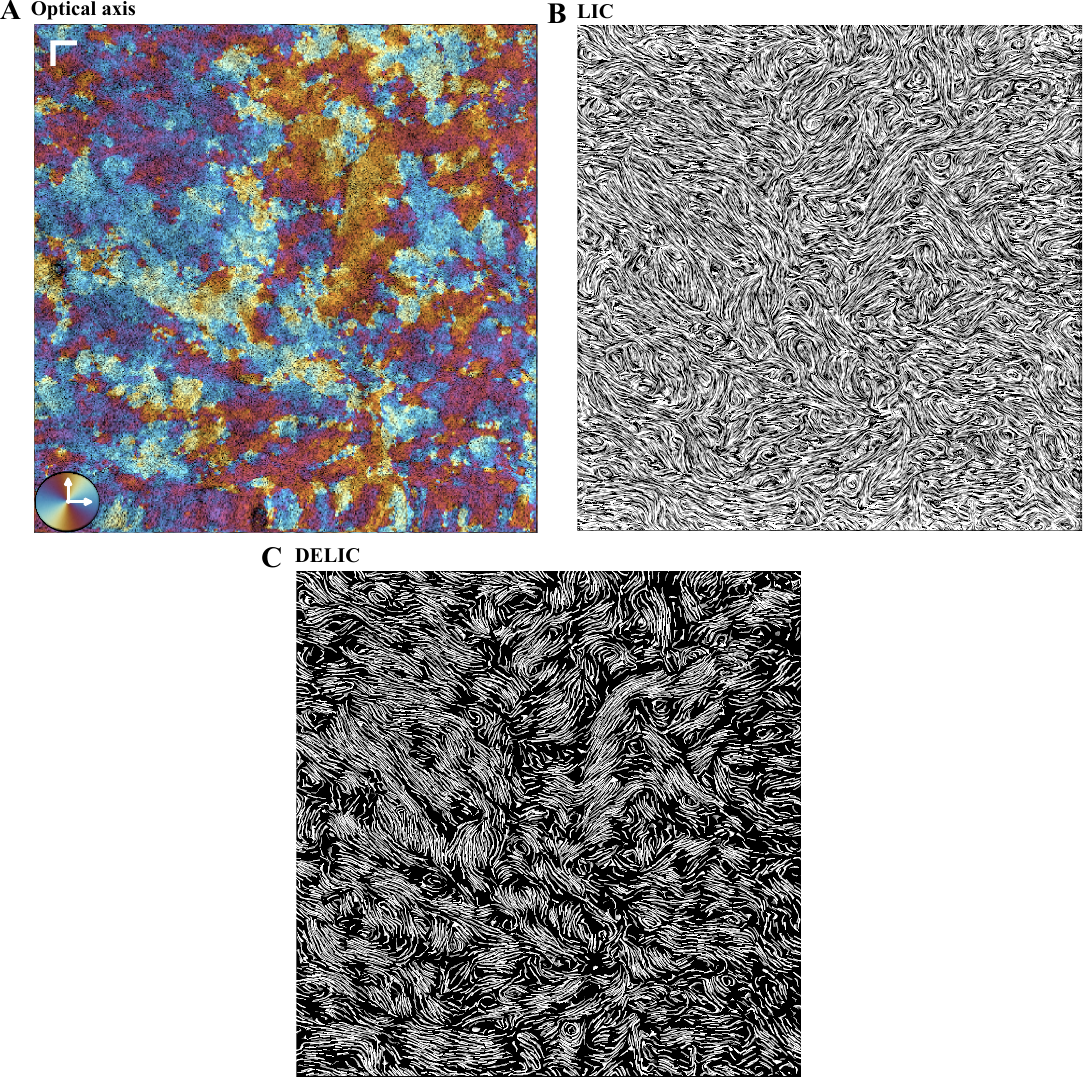}
    \caption{ (A) Optical axis image of human dorsal hand skin. Vertical and horizontal scale bars are 500~$\mu$m long. LIC (B) and DELIC (C) image of human skin. The DELIC image is generated using \(d_{\mbox{sep}}\) = 4 pixels within a domain of 512 by 512 pixels. Both LIC and DELIC utilised an integration step of 0.1 pixels and integration length of 50 steps.}
    \label{fig:results:Experimental}
\end{figure}

Fig.~\ref{fig:results:Experimental} highlights the distinct advantages of the DELIC method compared to conventional LIC when visualising biological tissue, particularly in capturing collagen fibril orientations within the skin. Although LIC (Fig.~\ref{fig:results:Experimental} B) successfully preserves the intricate details of the fibrillar structure, it tends to overwhelm the viewer with an overabundance of high-frequency information. This saturation makes it difficult to clearly interpret the directional patterns of the optical axis field. In contrast, the DELIC approach (Fig.~\ref{fig:results:Experimental} C) offers a much more digestible and coherent representation. By coarsening the output, DELIC offers a more interpretable visualisation where the orientation of the optical axis is immediately apparent.

A particular advantage of DELIC lies in its ability to encode physical tissue properties directly into the tractogram. The density of the rendered fibers corresponds to the sample's birefringence, establishing a clear physical connection between the tissue and the visualisation. This feature makes the tractograms more meaningful from a biophysical perspective, allowing users to intuitively grasp the variations in tissue birefringence.

\subsection{DELIC limitations and strengths}

While DELIC presents numerous advantages, it also entails some trade-offs in processing time. The primary bottleneck arises from the computation of the Voronoi tesselation during each iteration of Lloyd's algorithm. Current algorithms for Voronoi tessellation are limited in efficiency, with no known method that surpasses \(O(n \log n)\) time complexity in 2D~\cite{voronoi_nlogn} and even less efficient options available for 3D implementations. Exploring alternative methods for computing the CVT that avoid explicit Voronoi tessellation computations could significantly improve processing efficiency, particularly for 3D tractography applications. Though, we have found Lloyd's algorithm to be sufficiently fast in 2D.

Although this study emphasises regions of high birefringence, this assumption may not universally apply. The current implementation utilises birefringence strength to dictate visualisation density; however, DELIC is inherently flexible by design and can easily accommodate various encoding fields. Users could opt to prioritise areas where the optical axis exhibits rapid changes, improving density in these regions. This flexibility allows DELIC to be tailored to diverse research and clinical applications.

Moreover, DELIC's principles extend beyond polarimetry-based imaging. Its framework can be applied effectively in fields such as fluid dynamics, where velocity-encoded tractograms can visualise complex flow fields. By tailoring the encoding field, DELIC can emphasize critical features in different types of vector fields, underscoring its potential for broader applicability beyond biomedical tractography. This adaptability positions DELIC as a foundational tool for establishing a general framework, encouraging researchers to build upon it by developing custom encoding fields for their specific visualisations.

Ultimately, DELIC not only enhances our understanding of biological structures but also paves the way for innovative applications in diverse scientific domains. 


\section{Conclusion}

In conclusion, the density-encoded line integral convolution algorithm marks a significant advancement in visualising birefringent biological tissues. By employing a sparse noise texture and modulating tractogram density based on tissue birefringence strength, DELIC enhances the clarity and interpretability of collagen fibril orientations in birefringent tissue. Compared to traditional LIC, DELIC minimises clutter while retaining essential structural information, allowing researchers to focus on key areas without distraction. Although DELIC demands more computational resources than traditional methods, the increase in processing time remains relatively minimal, and the resulting enhancement in visualisation quality justifies this trade-off.

Ultimately, our innovative approach establishes a flexible framework that can adapt to various encoding fields, making it applicable across multiple scientific domains.

\section*{Acknowledgments}
 Funding was obtained through the Marsden Fund from the Royal Society Te Apārangi (UoA2110). The authors acknowledge Dr. Michael Hackmann, Dr. Quingyun Li and Dr. Boy Braaf for insightful discussions that aided in the data processing used to acquire polarimetry data. We also acknowledge the University of Auckland Doctoral Scholarship, which provided support for Darven Murali Tharan. Additionally, we acknowledge Tillmann Spellauge for his assistance with proofreading the manuscript.

\bibliography{Biblio} 

\begin{thebibliography}{10}
\providecommand{\url}[1]{#1}
\csname url@samestyle\endcsname
\providecommand{\newblock}{\relax}
\providecommand{\bibinfo}[2]{#2}
\providecommand{\BIBentrySTDinterwordspacing}{\spaceskip=0pt\relax}
\providecommand{\BIBentryALTinterwordstretchfactor}{4}
\providecommand{\BIBentryALTinterwordspacing}{\spaceskip=\fontdimen2\font plus
\BIBentryALTinterwordstretchfactor\fontdimen3\font minus \fontdimen4\font\relax}
\providecommand{\BIBforeignlanguage}[2]{{%
\expandafter\ifx\csname l@#1\endcsname\relax
\typeout{** WARNING: IEEEtran.bst: No hyphenation pattern has been}%
\typeout{** loaded for the language `#1'. Using the pattern for}%
\typeout{** the default language instead.}%
\else
\language=\csname l@#1\endcsname
\fi
#2}}
\providecommand{\BIBdecl}{\relax}
\BIBdecl

\bibitem{ghosh2011tissue}
N.~Ghosh and I.~A. Vitkin, ``Tissue polarimetry: concepts, challenges, applications, and outlook,'' \emph{Journal of biomedical optics}, vol.~16, no.~11, pp. 110\,801--110\,801, 2011.

\bibitem{mollink2017evaluating}
J.~Mollink, M.~Kleinnijenhuis, A.-M. v.~C. van Walsum, S.~N. Sotiropoulos, M.~Cottaar, C.~Mirfin, M.~P. Heinrich, M.~Jenkinson, M.~Pallebage-Gamarallage, O.~Ansorge \emph{et~al.}, ``Evaluating fibre orientation dispersion in white matter: Comparison of diffusion mri, histology and polarized light imaging,'' \emph{Neuroimage}, vol. 157, pp. 561--574, 2017.

\bibitem{yao2020high}
G.~Yao and D.~Duan, ``High-resolution 3d tractography of fibrous tissue based on polarization-sensitive optical coherence tomography,'' \emph{Experimental Biology and Medicine}, vol. 245, no.~4, pp. 273--281, 2020.

\bibitem{palm2010towards}
C.~Palm, M.~Axer, D.~Gr{\"a}{\ss}el, J.~Dammers, J.~Lindemeyer, K.~Zilles, U.~Pietrzyk, and K.~Amunts, ``Towards ultra-high resolution fibre tract mapping of the human brain-registration of polarised light images and reorientation of fibre vectors,'' \emph{Frontiers in human neuroscience}, vol.~4, p. 891, 2010.

\bibitem{tharan2025ConfProceedings}
D.~Murali~Tharan, M.~Bonesi, D.~Everett, M.~Goodwin, C.~McGoverin, S.~McGlashan, A.~Thambyah, and F.~Vanholsbeeck, ``Towards improved visualisation of polarisation-sensitive optical coherence tomography optical axis fields,'' in \emph{Polarized Light and Optical Angular Momentum for Biomedical Diagnostics 2025}, vol. 13322.\hskip 1em plus 0.5em minus 0.4em\relax SPIE, 2025, pp. 11--16.

\bibitem{lic}
B.~Cabral and L.~C. Leedom, ``Imaging vector fields using line integral convolution,'' in \emph{Proceedings of the 20th annual conference on Computer graphics and interactive techniques}, 1993, pp. 263--270.

\bibitem{olic}
R.~Wegenkittl, E.~Groller, and W.~Purgathofer, ``Animating flow fields: rendering of oriented line integral convolution,'' in \emph{Proceedings. Computer Animation'97 (Cat. No. 97TB100120)}.\hskip 1em plus 0.5em minus 0.4em\relax IEEE, 1997, pp. 15--21.

\bibitem{drexler2008optical}
W.~Drexler and J.~G. Fujimoto, \emph{Optical coherence tomography: technology and applications}.\hskip 1em plus 0.5em minus 0.4em\relax Springer Science \& Business Media, 2008.

\bibitem{psoct}
J.~F. De~Boer, C.~K. Hitzenberger, and Y.~Yasuno, ``Polarization sensitive optical coherence tomography--a review,'' \emph{Biomedical optics express}, vol.~8, no.~3, pp. 1838--1873, 2017.

\bibitem{baumann2017polarization}
B.~Baumann, ``Polarization sensitive optical coherence tomography: a review of technology and applications,'' \emph{Applied Sciences}, vol.~7, no.~5, p. 474, 2017.

\bibitem{sobol}
P.~Bratley and B.~L. Fox, ``Algorithm 659: Implementing sobol's quasirandom sequence generator,'' \emph{ACM Transactions on Mathematical Software (TOMS)}, vol.~14, no.~1, pp. 88--100, 1988.

\bibitem{cvt}
Q.~Du, V.~Faber, and M.~Gunzburger, ``Centroidal voronoi tessellations: Applications and algorithms,'' \emph{SIAM review}, vol.~41, no.~4, pp. 637--676, 1999.

\bibitem{lloyds}
Y.~Lu and H.~H. Zhou, ``Statistical and computational guarantees of lloyd's algorithm and its variants,'' \emph{arXiv preprint arXiv:1612.02099}, 2016.

\bibitem{li2018robust}
Q.~Li, K.~Karnowski, P.~B. Noble, A.~Cairncross, A.~James, M.~Villiger, and D.~D. Sampson, ``Robust reconstruction of local optic axis orientation with fiber-based polarization-sensitive optical coherence tomography,'' \emph{Biomedical optics express}, vol.~9, no.~11, pp. 5437--5455, 2018.

\bibitem{villiger2016deep}
M.~Villiger, D.~Lorenser, R.~A. McLaughlin, B.~C. Quirk, R.~W. Kirk, B.~E. Bouma, and D.~D. Sampson, ``Deep tissue volume imaging of birefringence through fibre-optic needle probes for the delineation of breast tumour,'' \emph{Scientific reports}, vol.~6, no.~1, p. 28771, 2016.

\bibitem{li2020vivo}
Q.~Li, D.~D. Sampson, and M.~Villiger, ``In vivo imaging of the depth-resolved optic axis of birefringence in human skin,'' \emph{Optics letters}, vol.~45, no.~17, pp. 4919--4922, 2020.

\bibitem{voronoi_nlogn}
L.~P. Chew and S.~Fortune, ``Sorting helps for voronoi diagrams,'' \emph{Algorithmica}, vol.~18, no.~2, pp. 217--228, 1997.

\end{thebibliography}
\bibliographystyle{IEEEtran}
\end{document}